\documentclass[12pt,letterpaper]{iopart}
\catcode`\*=11
\makeatletter
\let\oldequation*\equation*
\let\equation*\@undefined
\let\oldendequation*\endequation*
\let\endequation*\@undefined
\makeatother
\catcode`\*=12

\usepackage{amsmath} 
\usepackage{amssymb}
\usepackage{color}
\usepackage{graphicx}
\usepackage{mathtools}
\usepackage{lipsum}
\usepackage[square,sort&compress,numbers]{natbib}
 
\usepackage{enumitem} 

\begin{document}
\title{Running Measurement Protocol for the Quantum first-detection problem}
%\title{Running Measurement Protocol: The interface between quantum and classical dynamics}
\author{Dror Meidan$^{1,2}$, Eli Barkai$^{1,2}$ and David A. Kessler$^1$}
\address{$^1$Department of Physics, Bar-Ilan University, Ramat-Gan 52900, Israel}
%\author{Eli Barkai}
\address{$^2$Department of Physics, Institute of Nanotechnology and Advanced Materials, Bar-Ilan University, Ramat-Gan 52900, Israel}
\begin{abstract}
    The problem of the detection statistics of a quantum walker has received increasing interest, connected as it is with to the problem of quantum search. We investigate the effect of employing a moving detector, using a projective measurement approach with fixed sampling time $\tau$,  with the detector moving right before every detection attempt. For a tight-binding quantum walk on the line, the moving detector allows one to target a specific range of group velocities of the walker, qualitatively modifying the behavior of the quantum first-detection probabilities. We map the
    problem to that of a stationary detector with a modified unitary evolution operator and  use established methods for the solution of that problem to study the first-detection statistics for a moving detector on a finite ring and on an infinite 1D lattice. 
    On the line, the system exhibits a dynamical phase transition at a critical value of $\tau$, from a state where detection  decreases exponentially in time and the total detection is very small, to a state with power-law decay and a significantly higher probability to detect the particle. The exponent describing the power-law decay of the detection probability at this critical $\tau$ is $10/3$, as opposed to $3$ for every larger $\tau$. In addition, the moving detector strongly modifies the  Zeno effect. 
\end{abstract}

\submitto{\jpa}
\emph{Special Issue on New Trends in First-Passage Methods and Applications in the Life Sciences and Engineering}
%\author{David A. Kessler}
%\affiliation{Department of Physics, Bar Ilan University, Ramat-Gan 52900, Israel}
\maketitle
\section{Introduction}
The problem of detecting a quantum walker has been attracting increasing interest, as many search and decision algorithms can be modeled as a continuous-time quantum walk~\cite{Aharonov}, as established by Farhi, et al.~\cite{Farhi}. (For a general review of quantum walks, see Refs. \cite{BlumenMulken,Venegas}). Recent experimental advances have made it possible to measure a quantum walk at the single particle level \cite{intro1,intro2,intro3}. In addition, the theoretical side of the quantum first-detection problem for both continuous and discrete time walks, and more generally, any unitary evolution of a quantum system, has been the focus of study as it deals with the basic issue of when a given target state will first be detected~\cite{Dhar1, Dhar2, Harel,infinite_line, spectral_dim}. In general, the system is defined as a graph and the Hamiltonian $\hat{H}$ includes both hopping and site energy terms. The  wave function, initially localized at node $|x_i\rangle$, is probed stroboscopically at a node $|x_d\rangle$ with period $\tau$. This problem, reminiscent of the well-known classical first passage time problem, shares some of its aspects but also exhibits striking qualitative differences.

In the quantum problem, the sampling time $\tau$, determined in principle by the experimenter, is crucial. Too small $\tau$ will lead to the Zeno effect, which implies the particle is not detected if the initial and detected state are orthogonal, while too large $\tau$ will make the whole approach impractical. Hence there exists an optimal $\tau$, which can be chosen either to maximize the total detection probability or the mean time to detection, conditioned on successful detection~\cite{Harel}.  

\begin{figure}[ht]
\includegraphics[width=9.5cm, height=6cm]{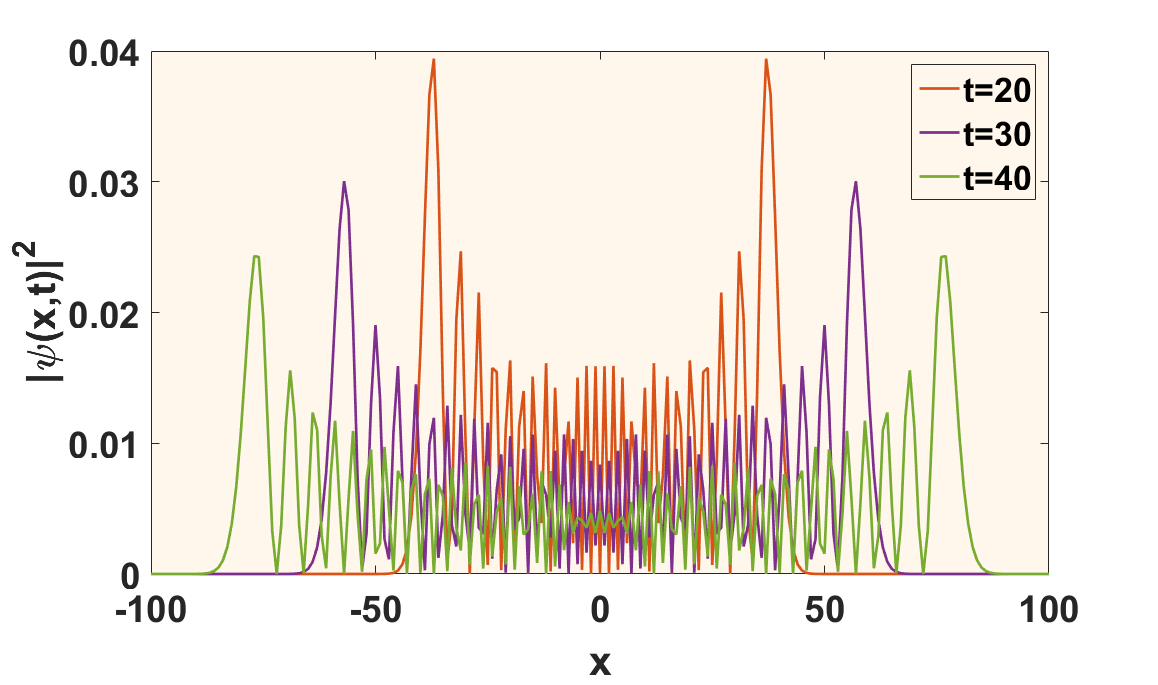}
\centering
\caption{The probability to measure the quantum walker without intermediate detection for several times ($t=20,30,40$) for the tight-binding model on a 1D lattice defined in Eq. \eqref{eq:Hinf}, with $\gamma=1$. The walker is initially  located at site $|0\rangle$. The time-dependent probability density, as well known, is given by $\rho(x,t)=|\psi(x,t)|^2=(J_{x}(2t))^2$ where $J_x$ is the  Bessel function of the first kind. In the interval between the peaks, $\rho(x,t)\sim {\cal{O}}(1/t)$ and outside, the spatial decay is exponential. Unlike a classical walk with its roughly Gaussian distribution, the most likely events are on the peaks spreading out ballistically, while at center destructive interference reduces the probability.}
\label{fig1}
\end{figure}

 As quantum particles move ballistically, and given the fact that the measurement is local, in an infinite system the particle will typically leave the stationary detector far behind, leading to a rapid falloff of the detection probability in time, compared to the classical case.  This ballistic motion \emph{in the absence of any measurement} is demonstrated in Fig. \ref{fig1}, for the case of a nearest-neighbor tight-binding Hamiltonian on the infinite line,
\begin{equation}
    \hat{H}=-\gamma\sum_k \big(|k \rangle\langle k+1|  + |k+1 \rangle\langle k|\big),
\label{eq:Hinf}
\end{equation}
where the spatial probability distribution, $\psi(x,t)|^2$, of a state initially localized at the origin is seen to exhibit two symmetrically located peaks, which move outward at constant velocity, the maximal group velocity of the state. The probabilities in the interior region decays in time as $1/t$, with characteristic quantum oscillations, and decays exponentially in space beyond the peaks. Actually, this is an upper bound; the \emph{first-detection probability} falls off faster than this, as $t^{-3}$, due to the cumulative effects of unsuccessful measurements~\cite{Harel}. It is thus interesting to  examine the possibility of  moving the detector after every  measurement attempt, so that the detection site  follows the walker along the graph. This Running Measurement Protocol allows one to maximize the total detection probability by adjusting the detector velocity (via the measurement period). 
\\\\ The questions to be tackled are as follows: Given $\tau$ and the tight-binding Hamiltonian on a ring or infinite line, what are the properties of the first-detection probability distribution at attempt $n$, which we denote $F_n$?  What is the asymptotic  behavior of $F_n$ for large $n$? In particular, since in the tight-binding model, the velocity of the walker is bounded, how does the asymptotic behavior change where the velocity of the detector is greater, slower or exactly at this bound? In addition, what is the behavior of $F_n$ when the detector traverses classically forbidden and allowed regions of the evolving particle? As demonstrated in previous studies, the particle is not always detected~\cite{krovi2006quantum,krovi2007quantum,Dhar1,Harel}, so what is the optimal sampling rate for which the total detection is maximized? How is the Zeno effect, discussed in Ref. \cite{Harel}, affected by the moving of the  detector? 

\section{The Running Measurement Protocol}
The detection protocol with a moving detector we study here (which we call the Running Measurement Protocol (RMP)), is a straightforward modification of the previous protocol we and others have studied for a stationary detector (which for convenience we refer to as the Stationary Measurement Protocol (SMP)).
As before, the system is represented by the Hilbert space $\cal{H}$. The wave function of the particle is initially prepared in the state $|\psi_{in}\rangle\in \cal{H}$. The particle dynamics is governed by  the Schr\"{o}dinger equation with a time-independent Hermitian Hamiltonian $\hat{H}$,  and a stroboscopic detection measurement is applied to the current state of the system with time period $\tau$, checking if the particle in the state $|\psi_d\rangle\in \cal{H}$. The two possible answers are 'yes' if the particle is detected in the probed location or 'no' otherwise. The experiment is terminated when we get a positive answer. The first attempt which yields a positive answer is denoted by $n$. %(Note that the first measurement is made at time $\tau$ after the initial preparation of the system, as opposed to at time 0 in the work, e.g., of Krovi and Brun~\cite{krovi2006quantum,krovi2007quantum}. In this latter protocol, there is no nontrivial SMP ``return" problem \cite{grunbaum2013recurrence}, where the initial and detection states are the same).
The statistics of this first-detection attempt $n$, and the elapsed time $n\tau$ are the focus of our interest. The new feature in our RMP is that immediately prior to performing the measurement, we move the detector right by one lattice spacing. In this paper we will discuss only  systems with translation symmetry, in particular focussing on the finite ring and the infinite line.

Mathematically, the strong stroboscopic detection (also known as von Neumann detection) is a projection onto $|\psi_d(n)\rangle$,
the current detection site, and so the measurement operator is represented as $\hat{D}(n)=|\psi_d(n)\rangle\langle\psi_d(n)|$. An unsuccessful detection causes a collapse of the particle's wave function, equivalent to acting with the operator $1-\hat{D}(n)$, and subsequent renormalization of the wave function. This is the collapse postulate \cite{cohen1977quantum}.  In addition, the unitary propagation operator of the quantum walker during the inter-measurement interval is $\Hat{U}(\tau)=e^{-i\Hat{H}\tau}$. 

The procedure of unitary evolution followed by strong measurement is repeated until the first detection is accomplished. This first-detection protocol combines the collapse of the wave function with the unitary dynamics generated by the Hamiltonian.

\begin{figure}[ht]
\includegraphics[width=8cm, height=4cm]{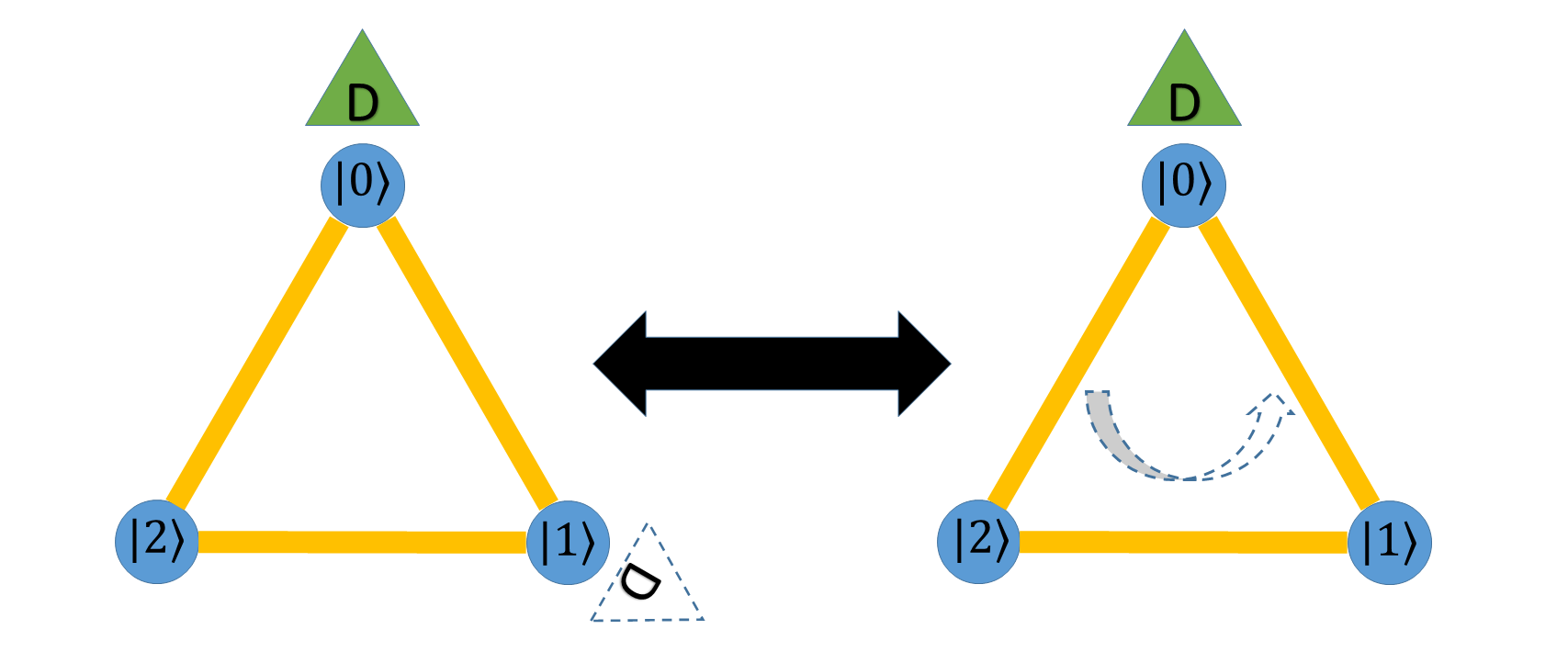}
\centering
\caption{This figure shows an example of the  equivalent Stationary Measurement Protocol problem for the Run Measurement Protocol on a ring graph with three sites. In the original system (the left illustration) the detector moves one site to the right before every measurement (i.e. every $\tau$ seconds), so the detection operator is $\hat{D}(n):=|\psi_{d}(n)\rangle\langle \psi_{d}(n)|=\hat{S}^n|\psi_{d}(0)\rangle\langle \psi_{d}(0)|(\hat{S}^\dagger)^n$, and the propagator of the particle is $\hat{U}(\tau)= e^{-i\hat{H}\tau}$.  In the mapped system (the right illustration), the propagator includes a rotation  of the  ring to the left (i.e. $\hat{S}^\dagger$) and the detection operator operates at a constant location, $D:=|\psi_d(0)\rangle\langle\psi_d(0)|$.  }
\label{fig: Pro RMR3}
\end{figure}

As shown in Ref. \cite{Dhar1,Harel}, the first-detection probability at attempt $n$, $F_n$, for the SMP is the square norm of the detection amplitude,
   $\varphi_n$, defined by:
\begin{equation}
    \varphi_n:=\langle\psi_d|\hat{U}(\tau)[(1-\Hat{D})\Hat{U}(\tau)]^{n-1}|\psi_{in}\rangle,
\end{equation} so
\begin{equation}
    F_n=|\varphi_n|^2.
    \label{eq: Fnvarphin}
\end{equation}
For our RMP protocol, this generalizes immediately to
\begin{equation}
    \varphi_n:=\langle\psi_d(n)|\hat{U}(\tau)\prod_{k=1}^{n-1} \left[(1-\Hat{D}(k))\Hat{U}(\tau)\right]|\psi_{in}\rangle .
\end{equation}
%To evaluate $\varphi_n$,
%we map our problem, assuming the Hamiltonian is translation invariant, to an equivalent stationary detector problem with a modified propagator 
%\begin{equation}
 %   \Tilde{U}(\tau)\equiv \Hat{S}^\dagger \Hat{U}(\tau)
%\end{equation}
To talk about the moving detector, we  define the shift operator $\Hat{S}$, where: 
\begin{equation}
\Hat{S}|x\rangle = |x+1\rangle .
\label{eq: S def}
\end{equation}
Then, in our problem,
\begin{equation}
   \hat{D}(k)=\hat{S}^k\hat{D}(0)(\hat{S}^\dagger)^k .
\end{equation}
An in this paper we assume that the Hamiltonian is translation invariant, we have $S^{\dagger}\Hat{U}(\tau)S=\Hat{U}(\tau)$. Then,
\begin{align}
    \varphi_n&=\langle\psi_d(n)|\hat{U}(\tau)\prod_{k=1}^{n-1} \left[(1-\Hat{D}(k))\Hat{U}(\tau)\right]|\psi_{in}\rangle\nonumber\\
    &= \langle\psi_d(0)|(S^{\dag})^n\hat{U}(\tau)\prod_{k=1}^{n-1}\left[ (\hat{S}^{\dag})^k(1-\Hat{D}(0))S^k\Hat{U}(\tau)\right]|\psi_{in}\rangle\nonumber\\
    &=\langle\psi_d(0)|\prod_{k=1}^{n-1}\left[ S^{\dagger}\Hat{U}(\tau) (1-\Hat{D}(0))\right]S^\dagger\Hat{U}(\tau)|\psi_{in}\rangle .
\end{align}
%which, as claimed above, is the SMP result with the propagator $S^\dagger \hat{U}(\tau)$.
 This equation leads to an  equivalence between the RMP and an effective SMP with a modified unitary operator, 
 \begin{equation}
\widetilde{U}=\hat{S}^\dagger e^{-i\hat{H}\tau}
\label{eq: Modifated Propagator}
\end{equation}
consisting of the standard propagator followed by a shift \emph{leftward}. It should be noted that the SMP with this propagator was previously considered by Sinkovicz, et al. \cite{sinkovicz2015quantized}, and here we see that the RMP naturally gives rise to this exact system.
The equivalence is generally applicable to systems with translation symmetry (such as our hopping Hamiltonian, Eq. \eqref{eq:Hinf}, on the infinite line or a finite ring) and says that moving the detector one site right is equivalent to moving the particle one site left. This argument  is shown for a ring with three sites in Fig. \ref{fig: Pro RMR3}. 

 For the RMP,  we define the ``return" problem as the case where the initial site of the detector and particle are equal (so that the first measurement occurs one site to the right of the initial site). Likewise, the ``arrival problem" describes the case when the detector and the particle start at different sites. Note that the first measurement takes place at time $\tau$, while other treatments \cite{krovi2006quantum,krovi2007quantum} take the first measurement to be at time $t=0$. This choice of protocol obviously yields very different physics when we consider the ``return"~\cite{grunbaum2013recurrence}  case.

Not all sequences of measurements lead to an eventual detection. This phenomenon is not unique for quantum systems; for example, it  also occurs for the classical random walk in three or higher dimensions. However, in the quantum world we may find non-detected dark states also in finite systems like a ring, unlike the classical counterpart~\cite{krovi2006quantum,krovi2007quantum,Dhar1,Harel}. We define the total detection probability:
\begin{equation}
    P_\textit{det}=\sum_{n=1}^\infty F_n.
    \label{eq: PdetDef}
\end{equation}
For numerical computations of $F_n$, given our mapping, it is convenient to use the  quantum renewal equation~\cite{Harel}:
\begin{equation}
    \varphi_n=\langle\psi_d|\widetilde{U}^{n}|\psi_{in}\rangle-\sum_{m=1}^{n-1} \langle\psi_d|\widetilde{U}^{n-m}|\psi_{d}\rangle\varphi_m.
\end{equation}
This equation connects the first-detection amplitudes with the free evolution of the wave function unperturbed by any measurement. It is a natural analogue of the well-known classical renewal equation~\cite{Redner}. The first term is the direct (i.e. absent of measurement) evolution from initial to detection state. The second term accounts for  the effects of previous unsuccessful detection measurements. 

The renewal equation is also the starting point for analysis of the properties of $F_n$, as it can be solved for $\varphi_n$~\cite{Harel} using the generating function approach of classical random walk theory. The $Z$ transform or discrete Laplace transform of $\varphi_n$ is by definition: 
\begin{equation}
    \varphi(z):=\sum_{n=1}^{\infty} z^n\varphi_n,
\end{equation}
(here we define $\varphi_0:= 0$, since the first-detection attempt happens at $t=\tau$),
and can be shown \cite{Harel} to be given by
\begin{equation}
    \varphi(z)=\frac{ \langle\psi_d|\widetilde{U}(z)|\psi_{in}\rangle - \langle\psi_d|\psi_{in}\rangle }{\langle\psi_d|\widetilde{U}(z)|\psi_d\rangle},
\end{equation}
where
\begin{equation}
    \widetilde{U}(z):=\sum_{n=0}^{\infty} z^n\widetilde{U}^{n}.
\end{equation}
The detection amplitude  $\varphi_n$ can then be recovered via a contour integral:
\begin{equation}
    \varphi_n=\frac{1}{2\pi i}\oint_{|z|=r}\frac{dz}{z^{n+1}}\, \varphi(z) .
    \label{eq: phiztophin}
\end{equation}
Note that $\varphi(z)$ is by definition analytic inside the unit disk, so the integration over the circle contour with radius $r\le 1$ only contains the one pole at the origin. In the ``return" problem,  we take $|\psi_d\rangle=|\psi_{in}\rangle$, and this equation simplifies to \cite{grunbaum2013recurrence}
\begin{equation}
    \varphi(z)= 1 - \frac{ 1}{\langle\psi_d|\widetilde{U}(z)|\psi_d\rangle}.
    \label{eq: REZRA}
\end{equation}
In the ``arrival" problem (i.e. $|\psi_d\rangle\neq|\psi_{in}\rangle$), we have, assuming that $|\psi_{in}\rangle$ and  $|\psi_{d}\rangle$ are orthogonal \cite{Harel}:
\begin{equation}
    \varphi(z)= \frac{\langle\psi_d|\widetilde{U}(z)|\psi_{in}\rangle}{\langle\psi_d|\widetilde{U}(z)|\psi_d\rangle}.
\end{equation}

Finally, we define some useful notation:
\begin{equation}
    u_n:=\langle\psi_d|\widetilde{U}^{n}|\psi_d\rangle; \qquad v_n:=\langle\psi_d|\widetilde{U}^{n}|\psi_{in}\rangle,
\end{equation}
and
\begin{align}
    u(z)&:=\sum_{n=0}^\infty u_n z^n=\langle\psi_d|\widetilde{U}(z)|\psi_d\rangle, \nonumber\\
    v(z)&:=\sum_{n=0}^\infty u_n z^n=\langle\psi_d|\widetilde{U}(z)|\psi_{in}\rangle.
\end{align}
\\\\The motion of the detector is thus seen to be equivalent to a stationary detector with a modified dynamics.  In particular, the new  dispersion relation depends on $\tau$, which gives rise to new behaviors. In the next two sections, we will encounter these new behaviors as we analyze the RMP on two simple graphs, the finite ring and the infinite line.

\section{The Ring}
In this section, we discuss the RMP on a ring graph with $L$ sites, with the tight-binding Hamiltonian 
\begin{equation}
    \hat{H}=-\gamma \sum_{j=0}^{L-1}\big(|j\rangle \langle j+1| + |j+1\rangle \langle j|\big),
\end{equation}
and cyclic boundary conditions, $|L\rangle=|0\rangle$.
The first-detection problem for this system with the SMP 
was discussed  in Ref. \cite{Harel}.   The eigenvalues and eigenvectors of $\hat{H}$ are: $E_k=-2\gamma\cos{\frac{2\pi k}{L}}$ and $|\chi_k\rangle$ defined with $\langle l|\chi _k\rangle=\frac{1}{\sqrt{L}} e^{-\frac{2\pi ikl}{L}}$, $k=0..{L-1}$. As noted above,
the shift operator $\hat{S}$ and the Hamiltonian $\hat{H}$ commute. The Hamiltonian is constructed from the shift operator $\hat{S}$ and its conjugate,    $\Hat{H}=\Hat{S}+\Hat{S}^\dagger$, so it has the same diagonal basis, namely the Fourier modes. The form of the shift operator in Fourier space is
\begin{equation}
\hat{S}=\sum_{k} e^{\frac{-2\pi i k}{L}} |\chi_k\rangle\langle \chi_k| .
\end{equation}
 The eigenvalues of the propagator $\hat{U}(\tau)$, are, by unitarity,  pure phases. We call these phases the dynamical phases of the system, (labelled by Fourier number $k$, and denoted by $\lambda_k $). One of the main properties of these dynamical phases is that they determine the exceptional $\tau$ values for which the moments of the first-detection time are discontinuous \cite{grunbaum2013recurrence,Harel}. The dynamic phases of the mapped RMP system  on the ring graph are:
\begin{equation}
    \lambda_k= -\frac{2\pi}{L} k- 2\gamma \tau\cos{\frac{2\pi k }{L} } .
    \label{eq: DP of RMP Ring}
\end{equation}

As shown by Gr\"unbaum, et al. \cite{grunbaum2013recurrence}, the most striking features of the ``return" problem in the SMP for a system with a discrete spectrum are that: 1) $P_\textit{det}=1$; and 2) $\langle n\rangle$ is equal to the number of unique dynamical phases (with non-zero overlap of the corresponding eigenstate with the detected state). Given our mapping, these results carry over immediately to the RMP case. In particular, for the ring of length $L$, the
number of unique dynamical phases is $L$, as opposed to the SMP where the number is roughly half this, namely $L/2+1$ for $L$ even, and $(L+1)/2$ for $L$ odd \cite{Harel}. This is due to the breaking of  reflection symmetry in the modified propagator in the equivalent SMP problem~\cite{sinkovicz2015quantized}.  At the exceptional values of $\tau$~\cite{Harel},  for which accidental degeneracies exist, the number of unique dynamical phases is reduced, and consequently, so is  $\langle n\rangle$, for both the SMP and RMP.

\begin{figure*}[t]
    \centering
    \includegraphics[width=17cm, height= 5cm]{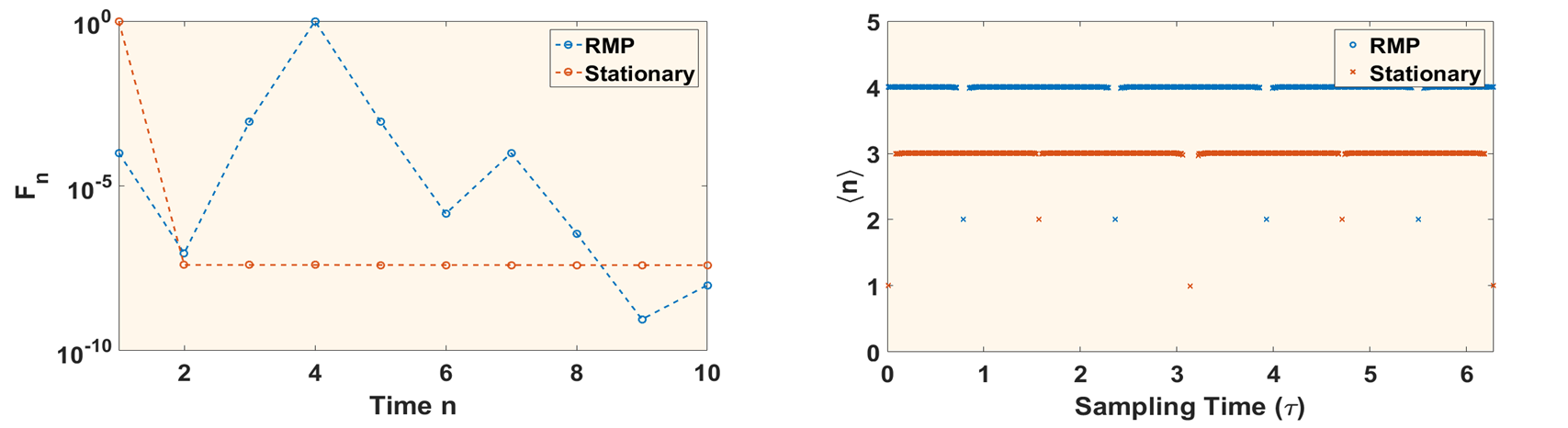}
    \caption{ Simulation results for the return problem for the SMP and RMP on a ring graph with four sites, $\gamma=1$. In the left panel, we plot the first-detection probability ($F_n$) with very small sampling time ($\tau=0.01$). For the SMP, the Zeno effect implies that for $n=1$ the probability to detect the particle is close to one, still as shown in the left panel $\langle n\rangle=3$. For the RMP, on the other hand, the Zeno effect is absent, but  there is a peak at $n=L$ instead. In the right panel, we present the average number of measurements till first detection,  $\langle n\rangle:=\sum_{k=1}^{\infty} kF_{k}/\sum_{k=1}^{\infty} F_{k}$, as a function of the sampling period $\tau$. Note that the Zeno effect makes $\tau=0$  an exceptional point for the SMP and $\langle n \rangle$ is discontinuous there, whereas it is smooth for the RMP. Due to the symmetry of the ring, the results of independent of the initial (= detection) site.}
    \label{fig: Ring4RP}
\end{figure*}

These results for  $\langle n\rangle$ are exemplified in the right panel of Fig.~\ref{fig: Ring4RP}, where we see that  $\langle n\rangle$ are 3 and 4 for the SMP and RMP cases respectively, barring the exceptional $\tau$. We see that, in this case, the RMP performs worse  than the SMP in the sense that the average return time is increased, and this is because in the RMP we break the reflection symmetry and so increase the number of unique phases, thus increasing $\langle n \rangle $, according to the general theorem of Gr\"unbaum, et al \cite{grunbaum2013recurrence}. We will soon show that the return problem is not typical in this regard.

\begin{figure*}[t]
    \centering
    \includegraphics[width=16cm, height=12cm]{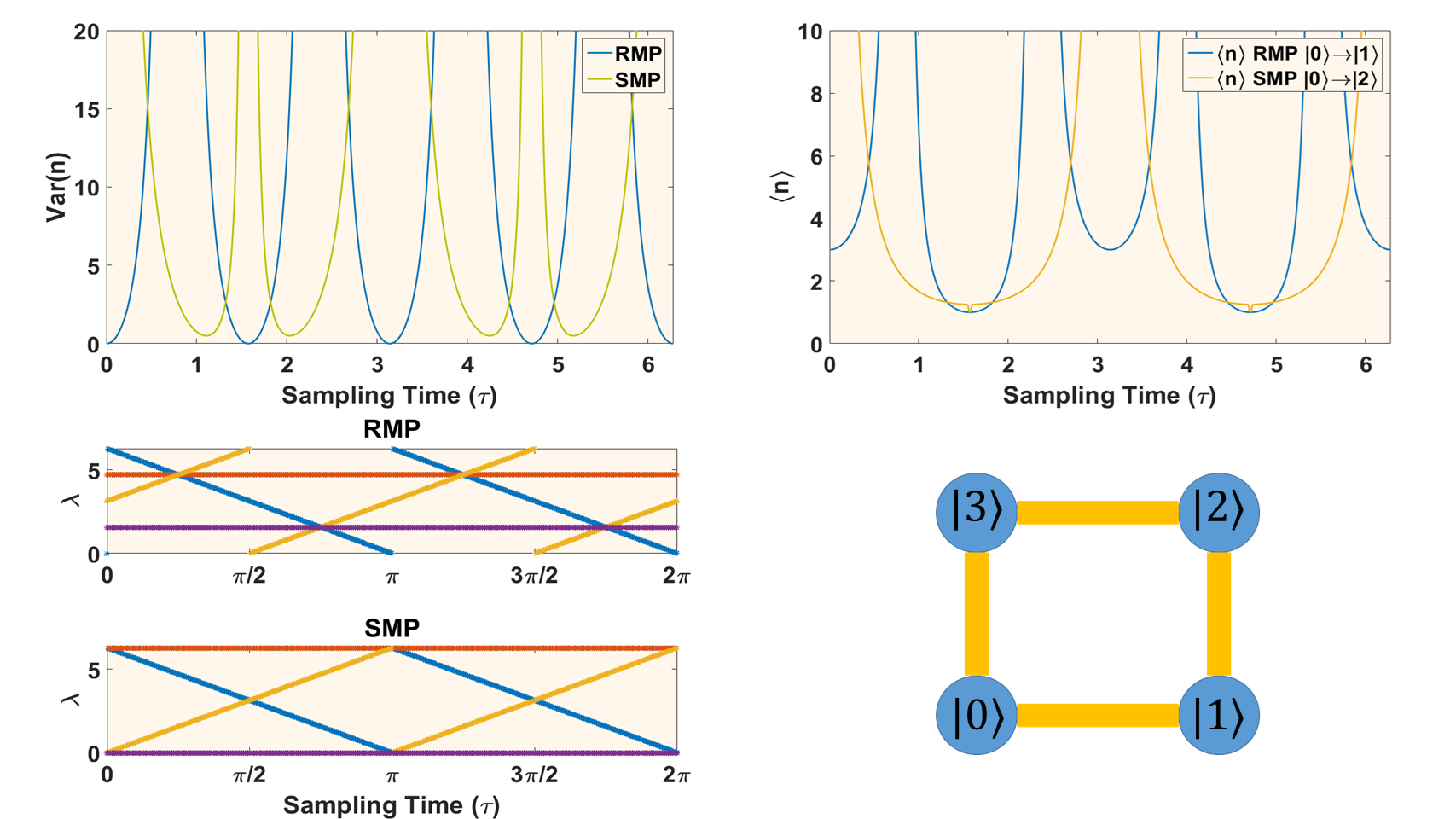}
    \caption{Numerical results for a quantum walk  with RMP on a ring with four sites  for the $|0\rangle\rightarrow|0\rangle$ ``return" problem (upper left) and the $|2\rangle\rightarrow|0\rangle$ "arrival" problem (upper right), $\gamma=1$. The upper left figure presents  the variance of the number of attempts to detect the particle vs $\tau$ in return problem. The upper right figure is the  average of the  number of attempts to detect the particle, approximated as  $\langle n\rangle=\frac{\sum_{k=1}^{N} kF_k}{\sum_{k=1}^{N} F_k}$ where $N=1000$, vs $\tau$, the results converging as we increase $N$ further (note that $P_{det}=1$ for these cases). The left bottom figure shows a plot of all the dynamical phases of RMP and SMP. The RMP dynamical phases are given in Eq. \eqref{eq: DP of RMP Ring}, the SMP phases are missing the first term. The figure shows the existence of exceptional (in $\langle n\rangle$) and divergence (in $\textit{Var}(n)$) points associated with each degeneracy of the dynamical phases. In addition, the reduction of variance at small $\tau$ in RMP case is clearly visible. Parenthetically, we point out  that at the exceptional case of $\tau=\pi/2$ for the SMP $|0\rangle \rightarrow |2\rangle$ arrival problem, $\langle n\rangle=1$, due to the fact that $U(\pi/2)$ exactly maps $|0\rangle$ to $|2\rangle$, leading to immediate detection, which happens as well for the RMP $|0\rangle\rightarrow|1\rangle$ arrival problem.}
    \label{fig:RMRingL4}
\end{figure*}

The difference in the $n$ dependence of $F_n$ in the return problem is even more striking.  This is especially true at small $\tau$, where the stationary detector protocol is dominated by the Zeno effect. There, the walker is almost surely detected on the first attempt, since it has not had time to move away from its initial position in the small time interval $\tau$. If, however, the walker was not detected on this first attempt, the chance to detect it on the subsequent attempt is very small, of order $\tau^4$, since the first failed measurement attempt has zeroed out the probability to be at the initial site and there is insufficient time to reconstitute its presence there.  This persists for subsequent measurements, and it takes of order $1/\tau^2$ attempts to accumulate the missing $O(\tau^2)$ total detection probability, as $F_n$ decays exponentially with $n$ at a rate proportional to $\tau^2$.  For the RMP, however, the picture is very different.  The fact that the detector moves makes the zeroing out of the probability at the detection site much less important. Thus, the first-detection is attempted one site distant from the initial site, so the probability of success is small, of order $\tau^2$. This is the case for subsequent attempts as well, until the detector returns to the original position of the particle on the $L$'th attempt. Then the particle is almost surely detected.  Subsequently, the detection probability drops by an additional power of $\tau^2$ after each series of $L$ measurements. Thus, the variance of $n$ for small $\tau$ is dramatically reduced in the RMP,
as seen in Fig.~\ref{fig:RMRingL4}, despite the small increase in $\langle n \rangle$, seen in Fig. \ref{fig: Ring4RP} for the ``return" problem on a ring graph with $L=4$ sites.  We see the increase from 3 to 4 for the average $n$ for all nonexceptional values of $\tau$.  Secondly, the sets of exceptional $\tau$'s are different in the two cases, due to the difference in the dynamical phases.  In the first panel, we show the comparison of $F_n$ for the two protocols, showing the peak at $n=1$ and the very slow decay for $n\ge 2$ for the stationary detector, compared to the peak at $n=4$ and the very rapid subsequent decay for RMP. In these and all subsequent figures, we have taken $\gamma=1$.


\begin{figure}[ht]
\includegraphics[width=9.5cm, height=5cm]{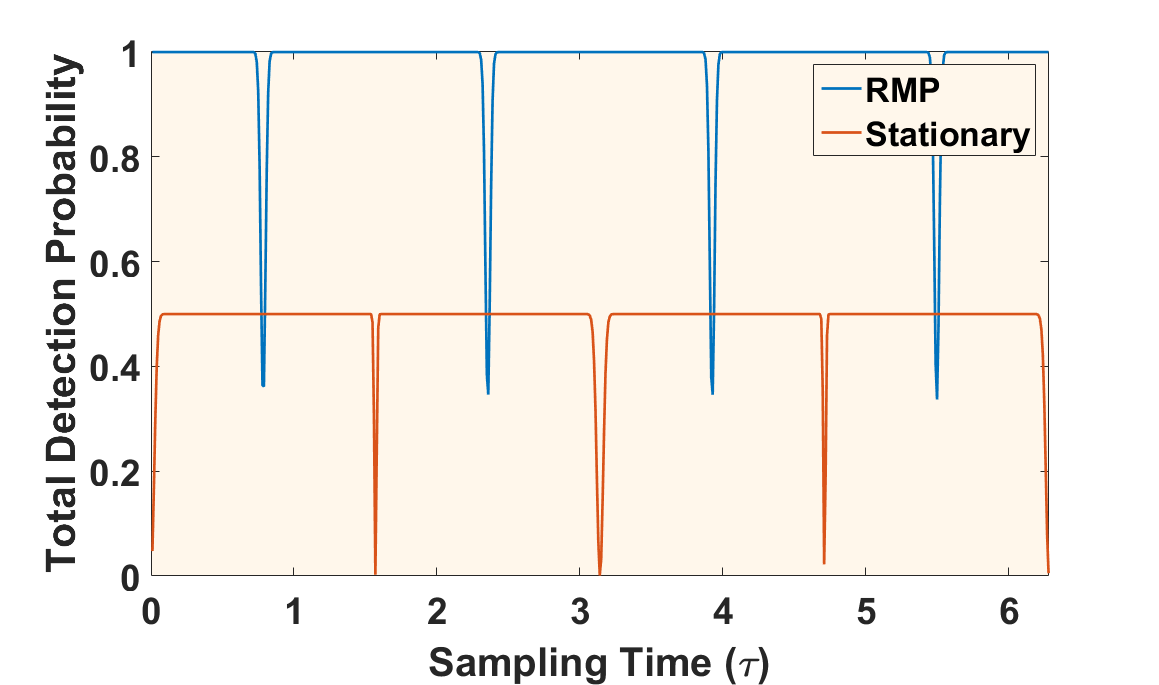}
\centering
\caption{Total detection probability for the SMP and RMP for the $|1\rangle\rightarrow|0\rangle$ arrival problem. For the SMP. the site $|1\rangle$ is typically half-dark because of the degeneracy of dynamical phases for all sampling time values. For specific $\tau$ values, the degeneracy is stronger and the total detection probability drops to zero. The RMP breaks the degeneracy of the dynamical phases so the the particle is almost surely detected.}
\label{fig: APPdet}
\end{figure}

The major difference in the arrival problem in the two problems is in $P_\textit{det}$.  For the SMP, $P_\textit{det}$ is typically less than unity for highly symmetrical graphs like the ring (this point is analysed in more detail  in~\cite{Inpre}), whereas for the RMP for nonexceptional $\tau$'s, it is unity, due to the nondegeneracy of the dynamical phases and in this sense, the RMP  performs better in this case.
This is seen in Fig.~\ref{fig: APPdet}, where we simulate the case of $|0\rangle\rightarrow|1\rangle$ for the SMP and RMP.   

\section{Infinite Line}
For the case of the moving detector on the infinite line with Hamiltonian: 
\begin{equation}
    \Hat{H}=-\gamma \sum_{j\in\mathbb{Z}}(|j\rangle \langle j+1| + |j+1\rangle \langle j|)
    \label{eq: ILH}
\end{equation}
Again we go to the moving frame using $\Hat{S}$, and the dynamical phase, which plays the role of $-\tau E$ in the SMP is
\begin{equation}
       \lambda(k)=  k + 2\gamma\tau\cos(k).
       \label{eq:phasek}
\end{equation}
 To obtain the last equation we took Eq. \eqref{eq: DP of RMP Ring}  and went  to the continuum limit with $L\rightarrow\infty$, so $k$ runs continuously in the range $-\pi < k \le \pi$. The behavior of $\lambda(k)$ changes at $\gamma\tau=1/2$. For $\gamma\tau<1/2$, corresponding to  detector motion with velocity $1/\tau$, which is faster than  the maximum group velocity of the particle, $2\gamma$, $\lambda(k)$ has no critical points satisfying $\lambda'(k)=0$, on the real interval $-\pi < k \le \pi$. On the other hand, for $\gamma\tau>1/2$, slow detector motion, there are two real critical points, which approach the values of $0$ and $\pi$ for $\tau\to\infty$, i.e., with no detector motion. The connection between $\varphi_n$ and the dynamical phase relation (which for SMP is just the dispersion relation) in the return  problem was discussed in Ref. \cite{spectral_dim}. As we shall see, the conversion from $\varphi(z)$ to $\varphi_n$ yields qualitatively different results in the fast and slow detector cases. We use the same contour integral procedure used for the stationary detector in \cite{infinite_line}.

\subsection{Return problem, fast-moving detector case}
We first treat the fast detector case, $\tau<1/2$, starting on $|0\rangle$ and measuring there.  We need the singularity structure of  $u(z)=\langle0|\widetilde{U}(z)|0\rangle=  \sum_{n=1}^\infty u_n z^n$, where $u_n=\langle0|\widetilde{U}^n|0\rangle$, which is determined by the large-$n$ behavior of $u_n$. For the moving detector, we have
\begin{equation}
u_n=\frac{1}{2\pi}\int_{-\pi}^{\pi} dk e^{-in\tau(-\frac{k}{\tau}-2\gamma\cos(k))}=(i)^n J_n(2n\gamma\tau),
\label{eq. unRMIL}
\end{equation}
as opposed to the same expression with $J_0$ for the stationary detector.

\begin{figure*}[t]
    \centering
    \includegraphics[width=16cm, height=12cm]{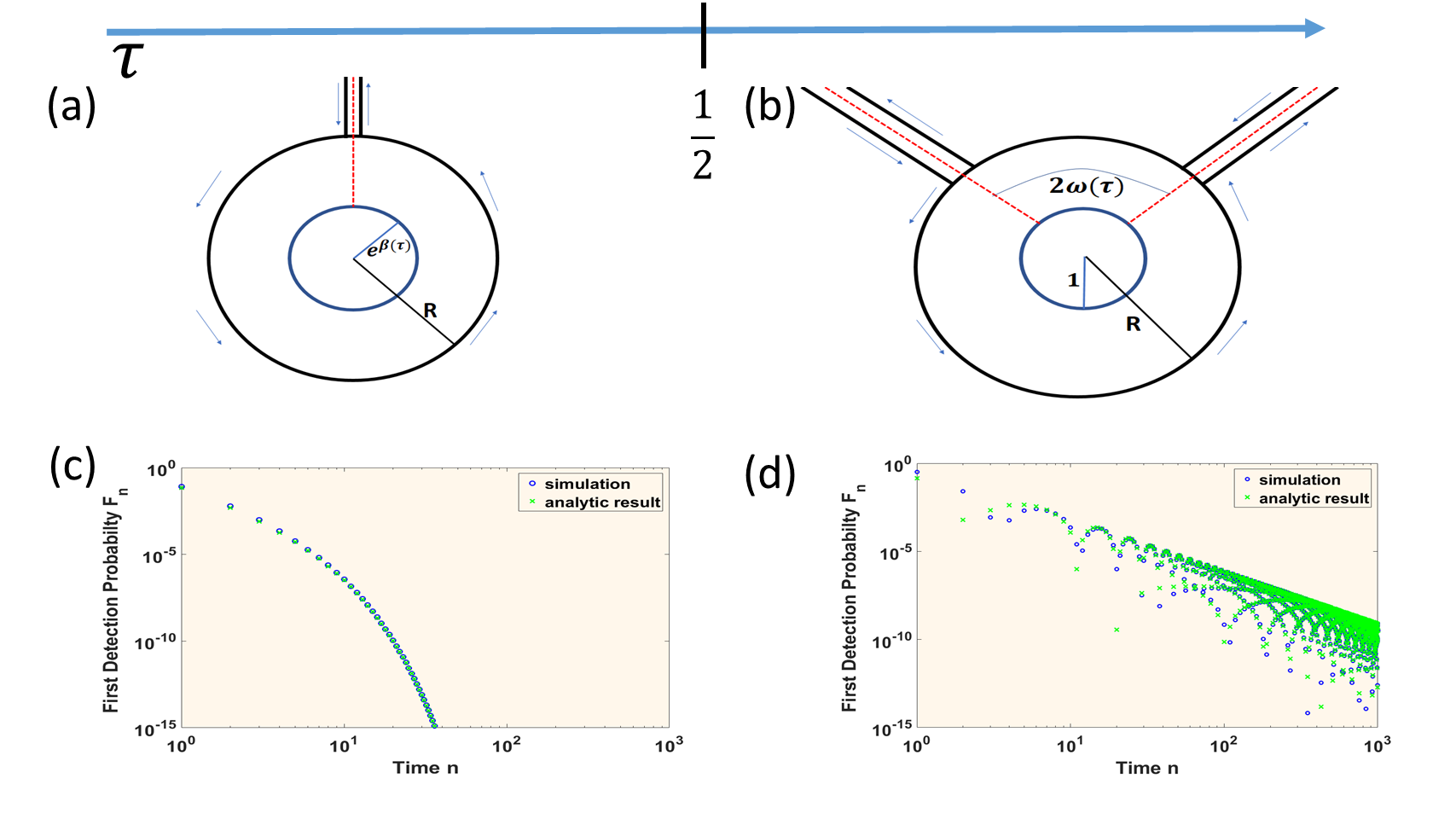}
    Return problem: \caption{a) Sketch of the integration contour for $\oint dz\,\frac{\varphi(z)}{z^{n+1}}$. For $\tau<1/2$, (with $\gamma=1$) the integrand has a branch-cut (red line) starting at the radius $e^{\beta(\tau)}$ going out to infinity along $z^{\star}(y)$. The radius $R$ of the contour is taken to infinity avoiding the cut. b) Sketch of the integration contour for $\oint dz\,\frac{\varphi(z)}{z^{n+1}}$. For $\tau>1/2$ the integrand has two branch-cuts (red line), starting on the unit circle and going out to infinity twice along $z_{-1}^{\star}(y)$ and $z_{1}^{\star}(y)$. The radius $R$ of the contour is taken to infinity except near the cuts, where the contour goes inward and wraps around the cuts. c-d) Probability of quantum first-detection, $F_n$ for the return  problem with the Run Measurement Protocol on an infinite line tight-binding model. $\gamma=1$. c) The fast-moving detector case, with $\tau=0.3$.  Blue circles are numerical results, red stars are the predictions of Eq. \eqref{eq: frfm}.   The exponential decay is clearly visible. d) The slow-moving detector case, with $\tau=0.8$.  Blue circles are numerical results, red stars are the predictions of Eq. \eqref{eq: frsm}.  The power-law decay and oscillations are clearly visible.}
    \label{fig: Fn&trajectory}
\end{figure*}

We can then use the approximation from \cite{G-R} (Eq.~8.452, p.~921): when $\nu\rightarrow\infty$ and the index of the Bessel function is greater than its argument, 
\begin{equation}
    J_{\nu}\left(\frac{\nu}{\cosh\alpha}\right)\sim\frac{e^{\nu( \tanh\alpha\,-\,\alpha)}}{\sqrt{2\nu\pi \tanh\alpha}}\ .
\end{equation}
In our case $\cosh\alpha=1/(2\gamma\tau)$, so $\tanh\alpha=\sqrt{1-4\gamma^2\tau^2}$, and
\begin{equation}
v_n \sim \frac{C(\gamma\tau)(-i)^n e^{-\beta(\gamma\tau)n}}{\sqrt{n}},
\end{equation}
where $C(x):=[2\pi (1-4x^2)^{1/2}]^{-1/2}$ and $\beta(x):= \textrm{arccosh}(1/2x)-\sqrt{1-4x^2}$. For $0<x<1/2$, $\beta(x)$ diverges logarithmically as $x\to 0^+$, $\beta(x)\sim -\ln x$, and decreases to 0 as $x\to 1/2$. Near the singularity of $v(z)$, the large-$n$ terms dominate, and we can calculate $v(z)$:
\begin{equation}
u(z)= \sum_{n=1}^{\infty} u_n z^n \sim C(\gamma\tau)\textrm{Li}_{\frac{1}{2}}(e^{-i\pi/2-\beta(\gamma\tau)}z),
\end{equation}
where $\textrm{Li}_{\nu}(z)=\sum_{n=0}^{\infty}\frac{z^n}{n^{\nu}}$ is the polylogarithm. The polylogarithm  has a branch-cut in the complex plane along the positive real axis for $|z|>1$, and near the branch-cut \cite{PL110}:
\begin{equation}
\textrm{Li}_{\frac{1}{2}}(e^y+i0^{\pm})\sim \pm i\sqrt{\frac{\pi}{y}}.
\end{equation}
In the fast-moving detector case,  the branch-cut is along the part of the positive imaginary axis where $|z|>e^{\beta(\gamma\tau)}$. Parameterizing  the branch-cut as $z^\star(y):=e^{i\frac{\pi}{2}+\beta(\gamma\tau)+y}$, we have:
\begin{equation}
\label{eq: vzsprp}
u(z)\sim \pm i C(\gamma\tau)\sqrt{\frac{\pi}{y}}.
\end{equation}
 Substituting this into Eq. \eqref{eq: REZRA} and then using the small $y$ expansion, we get:
\begin{equation}
\varphi(z)\mid_{z=z^\star(y)}\approx1\pm\frac{i}{C(\gamma\tau)}\sqrt{\frac{\pi}{y}} .
\end{equation}
Using Eq. \eqref{eq: phiztophin}, we can now recover $\varphi_n$ via a contour integral $\frac{\varphi(z)}{z^{n+1}}$, along a sufficiently small circle surrounding the origin.  We can then deform the contour to infinity, leaving only the branch-cut contribution. The trajectory is sketched in Fig.~\ref{fig: Fn&trajectory}(a).
Along the branch-cut, we change variables to $y$ as above. For large $n$, the integral is of Laplace type, dominated by the region close to the singularity, and  the  discontinuity of $\varphi(z)$ along the branch-cut, denoted  $Disc[\varphi(z^\star(y))]$, is approximately:
\begin{equation}
Disc[\varphi(z^\star(y))]\sim\frac{2i}{C(\gamma\tau)}\sqrt{\frac{y}{\pi}}.
\end{equation}
Doing the contour integral, we get
\begin{equation}
\varphi_n\sim\frac{1}{2\pi i}\int_0^\infty dy\, (e^{i(\pi/2+\beta(\gamma\tau))})^{-n}e^{-ny}Disc[\varphi(z^\star(y))] .
\end{equation}
The final result is:
\begin{equation}
\varphi_n\sim\frac{1}{2\pi C(\gamma\tau)\sqrt{n^3}}(e^{-\beta(\gamma\tau)-i\pi/2})^n, 
\end{equation}
and so
\begin{equation}
\label{eq: frfm}
F_n=|\varphi_n|^2 \sim \frac{\sqrt{1-4\gamma^2\tau^2}}{2\pi n^3}e^{-2n\beta(\gamma\tau)}.
\end{equation}
The analytic result matches the simulation results, as we see in Fig.~\ref{fig: Fn&trajectory}(c). While for the stationary detector, the power law $F_n\propto\frac{1}{n^3}$ was found previously, double the classical exponent $3/2$ from random walk theory \cite{Harel}, here the decay is exponential in $n$, modulated by the same exponent $3$.

In addition, because $\beta(x)$ is monotonically decreasing in the interval $0\le x\le \frac{1}{2}$ and $\lim_{x\rightarrow \frac{1}{2}}\beta(x)=0$,  the exponential decay is stronger for smaller $\gamma\tau$, and for  $\gamma\tau\ll 1$, $F_n$ attains its fastest decay rate, $F_n\sim (\gamma\tau)^{2n}$. In this ``Zeno" limit, $P_\textit{det}$ is dominated by $F_1$, and $P_\textit{det}\sim (\gamma\tau)^2$. In physical terms, when $\gamma\tau<1/2$ is small, the detector outruns the particle and so  the probability to see the particle  decays rapidly, an effect which gets stronger as the detector velocity $1/\tau$ increases.
\subsection{Return problem, slow detector case}
In this subsection, we treat the slow detector case, $\gamma\tau>1/2$. The calculation in principle proceeds as above, however here the argument of the Bessel function is greater than the index. The asymptotics of the Bessel function in this case are~\cite{G-R} (Eq.~8.453,p.~922)
\begin{equation}
    J_{\nu}\left(\frac{\nu}{cos(\beta)}\right)\sim\sqrt{\frac{2}{\nu\pi\tan(\beta)}}\cos\left(\nu\tan(\beta)-\nu\beta-\frac{\pi}{4}\right).
\end{equation}
Here, $\beta=\arccos(1/(2\gamma\tau))$ so $\tan(\beta)=\sqrt{4\gamma^2\tau^2-1}$, and the final approximation for large $n$ of $v_n$ is:
\begin{equation}
u_n\sim C(\gamma\tau)\frac{(-i)^n}{\sqrt{n}}\cos\left(\omega(\gamma\tau) n-\frac{\pi}{4}\right),
\end{equation}
where $C(x):=\sqrt{\frac{2}{\pi(4x^2-1)^{1/2}}}$ and $\omega(x):=\sqrt{4x^2-1}-\arccos(\frac{1}{2x})$. For $1/2<x<\infty$, $\omega(x)$ diverges linearly as $x\rightarrow\infty$ and decreases to 0 as $x\rightarrow 1/2^{+}$. Near the singularity of $u(z)$, again the large $n$ terms dominate, and so:
\begin{equation}
    u(z)
\sim\frac{C(\gamma\tau)}{2}\Bigg[e^{-i\frac{\pi}{4}}\textrm{Li}_{\frac{1}{2}}\left(e^{-i(\frac{\pi}{2}-\omega(\gamma
\tau)}\right)
+e^{i\frac{\pi}{4}}\textrm{Li}_{\frac{1}{2}}\left(e^{-i(\frac{\pi}{2}+\omega(\gamma\tau)}\right)\Bigg].
\end{equation}
In the slow-moving detector case, because of the phases in the arguments of the polylogarithm there are two line branch-cuts, at angles $\pm\omega(\gamma\tau)$ around the positive imaginary axis, with $1<|z|<\infty$. We parameterize the branch-cuts by $z_{\alpha}^\star(y):=e^{i\frac{\pi}{2}-i\omega(\gamma\tau)\alpha+y}$, where $\alpha=\pm 1$, and use the behavior of the polylogarithm near its singularity:
\begin{equation}
u(z)\sim\pm\frac{C(\gamma\tau)}{2}e^{i(\frac{\pi}{2}-\alpha\frac{\pi}{4})}\sqrt{\frac{\pi}{y}} .
\end{equation}
Substituting this into  $\varphi(z)=1-\frac{1}{u(z)}$ and expanding for small $y$, $\varphi(z)$ becomes:
\begin{equation}
\varphi(z)\mid_{z=z_\alpha^\star(y)}\approx1\pm i\frac{2}{C(\gamma\tau)}e^{i\frac{\pi}{4}\alpha}\sqrt{\frac{y}{\pi}} .
\end{equation}
Again, we can recover $\varphi_n$ via a contour integral deforming the contour to infinity, leaving only the two branch-cut contributions. The trajectory in Fig.~\ref{fig: Fn&trajectory}(b). Along the branch-cuts, we change variables to y as above. For large $n$, the integrals are again of Laplace type, dominated by the region close to the singularities, and the discontinuity of $\varphi(z)$ along the branch-cuts is approximately:
\begin{equation}
Disc[\varphi(z_{\alpha}^\star(y))]\sim\frac{4i}{C(\gamma\tau)}e^{i\frac{\pi}{4}\alpha}\sqrt{\frac{y}{\pi}} .
\end{equation}
Doing the contour integral, we get
\begin{equation}
\varphi_n\sim\sum_{\alpha}\frac{1}{2\pi i}\int_0^\infty dy(e^{i(\pi/2-\omega(\gamma\tau)\alpha)})^{-n}e^{-ny}Disc[\varphi(z^\star(y))] .
\end{equation}
The final result is:
\begin{equation}
\varphi_n\sim\frac{2e^{-i\frac{\pi}{2}n}\cos\left(\frac{\pi}{4}+n\omega(\gamma\tau)\right)}{\pi C(\gamma\tau)\sqrt{n^3}},
\end{equation}
and so:
\begin{equation}
\label{eq: frsm}
F_n\sim\frac{2\sqrt{4\gamma^2\tau^2-1}\cos^2\left(\frac{\pi}{4}+n\omega(\gamma\tau)\right)}{\pi n^3}
\end{equation}

This analytic result matches the simulation results, as we see in Fig.~\ref{fig: Fn&trajectory}(d). In addition, because $\omega(\gamma\tau)$ is monotonically increasing for $\gamma\tau>\frac{1}{2}$,  the oscillation frequency of $F_n$  increases with $\tau$. The frequency $\omega(\gamma\tau)$ has a supremum because the difference of phase which gave rise to $\omega(\tau)$ is limited by $2\pi$. This occurs at $\gamma\tau_c\approx 2.32$, where $\omega(\tau)<\pi$. In addition, we see the effect of the coinciding of the two branch-cuts, causing a change of  behavior of $P_{det}$ from monotonic increasing to oscillations.  It is obtained in the RMP return  and (as we shall see) arrival problems and also in the arrival problem on the infinite line with the SMP \cite{infinite_line}.    
\\\\We note that,  according to 
Eq.~\eqref{eq: frsm}, the leading-order large-$n$ behavior of $F_n$ vanishes in the  limit $\gamma\tau\rightarrow 1/2$.  This indicates that $F_n$ for $\gamma\tau=1/2$ behaves differently, and requires a more careful analysis,  to which we turn in the next subsection.

\subsection{Return problem: $\gamma\tau=0.5$}

The critical value $\gamma\tau=1/2$ marks the transition between the exponential decaying and oscillatory large-$n$ behavior of the first-detection probabilities, so we expect that the limit has only a pure  power-law decay without exponential or oscillatory behaviors. Physically, in this case, the detection site sees only the maximal group velocity of the particle and the participation of only one velocity eliminates the oscillations. In addition, the power-law of the decay is anomalous, due to the changing of the exponent of the power-law decay of $u_n$ from its typical value of $1/2$  to $1/3$,
as we shall see. More fundamentally, this will be seen to arise from  the emergence of an inflection point in the dynamical phase relation, Eq. \eqref{eq:phasek}. 

It was shown in Ref.~\cite{spectral_dim} that the power-law decay of $F_n$ is governed by the behavior of the dispersion relation at its extremum. For $\gamma\tau=1/2$, there is instead an inflection point, where both the first and second derivatives vanish, and the power-law is governed by the third order derivative.  The general relation between the behavior near the inflection point  and the order of power-law of $u_n$  will be discussed in Appendix B.    
\\\\In detail,  the calculation of $\varphi_n$  proceeds as above; therefore in this subsection, only the changes will be mentioned. First, here the argument of the Bessel function is equal to the index. The asymptotics of the Bessel function in this case are \cite{abramowitz1964handbook} (eq. 9.3.5 p.366): 
\begin{equation}
    J_n(n)\sim \frac{C}{n^{1/3}}
    \label{eq: AS 1}
\end{equation}
where $C=\frac{2^{1/3}}{3^{2/3}\Gamma(2/3)}=2^{1/3}\textrm{Ai}(0)$. $\textrm{Ai}(x)$ is the Airy function. The presence of the Airy function in this equation signals the transition between the oscillatory and exponential decaying regimes of $J_n$, as in the vicinity of the turning point in the Schr\"odinger Eq. 
\\\\ The different power-law in the large-$n$ asymptotics changes the index of polylogarithm in $u(z)$; this affects the behavior around its branch-cut. $u(z)$ has one branch-cut which runs from $z=-i$ to infinity on the negative imaginary axis. This implies that  $\varphi_n$ exhibits neither exponential decaying (because the branch-cut starts at $|z|=1$) nor oscillations (because there is only a single  branch-cut). Performing the integration around the branch-cut yields 
\begin{equation}
    F_n\sim\frac{\Gamma(-1/3)^2}{3\cdot 6^{2/3}\pi^2 n^{10/3}}
    \label{eq: FnGT0.5}
\end{equation}
The results of a simulation (not shown) of $F_n$ with the critical $\tau$  agree with this last equation.  
\subsection{Arrival Problem: Fast-moving detector}
\begin{figure*}[t]
\centering
\includegraphics[width=17cm, height= 5cm]{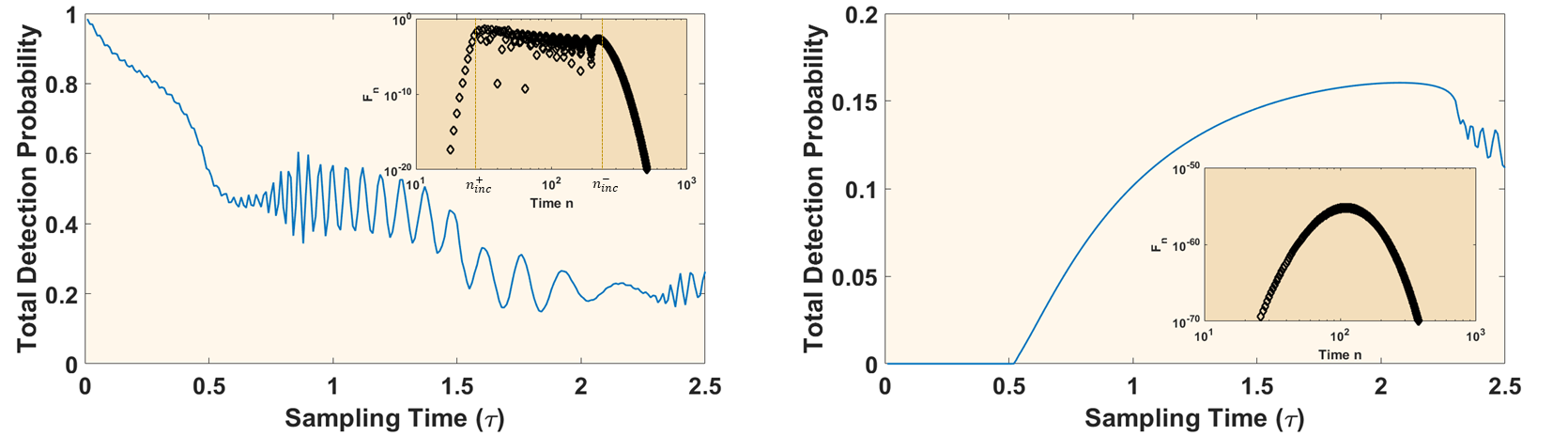}
\centering
\caption{The total detection probability Eq. \eqref{eq: PdetDef} for several cases of the RMP arrival problem for the tight-binding model on an infinite line. In the left figure, we present numerical results for $P_\textit{det}$ of the arrival problem with $\xi=-50$. In the inset, $F_n$ versus $n$ for $\tau=0.4$ is presented; the incidence times are $n_{inc}^{+}\approx 28$ and $n_{inc}^{-}\approx 250$; these results agree with Eq.~\eqref{eq: ncpm}. In the right figure, we present numerical results for $P_{det}$ of the arrival problem with $\xi=50$; notice the transition from a monotonic increase to oscillations at $\tau_c\approx 2.32$ because the discontinuities of the generating function ($\varphi(z)$) around this point. In the inset, $F_n$ versus $n$ for $\tau=0.4$ is shown, note the extremely small probabilities to detect the particle.  In both figures, the numerical $P_\textit{det}$ is the partial sum until $N=1000$. The results converge (not shown) as we increase $N$ further. In addition, the numerical results match the analytical results in subsection \ref{moveZeno}.}
\label{fig: PdetArrival}
\end{figure*}

We now turn to the arrival problem, where the experiment starts with the particle initially located at the origin, $|\psi_\textit{in}\rangle=|0\rangle$, and the detector at $x=\xi$, so that $|\psi_\textit{d}\rangle=|\xi\rangle$,
first looking at the fast-moving detector case. Similar to the return problem for the fast-moving detector, it can be shown that there is only one branch-cut in the arrival problem, simplifying the analysis of how changing the initial location of the detector  affects  the probabilities, as we show below in the case where the initial $\xi>0$. In addition, in the case of $\xi<0$, we will briefly explain the location of the critical points and the change of behavior at these points. The slow-moving detector case has two branch-cuts, so the analysis is similar to that of the stationary detector discussed in Ref.~\cite{infinite_line} and we will not revisit it here in this paper.   
\\\\When the initial location of the detector is to the left of the initial location of the particle, $\xi<0$, due to the fact that the fast-moving detector catches up with and passes the particle, there are two special times, marking the entrance ($n_{inc}^+$) and departure ($n_{inc}^-$) of the detector to the oscillatory zone between the peaks of the free particle wave-function (as shown in Fig. \ref{fig: PdetArrival}); note that in the slow-moving detector case there is only one $n$, which marks the entrance, and then we obtain the familiar behavior of power-law decay with oscillations. Before the entrance time, $F_n$ increases monotonically, then between  $n_{inc}^{+}$ and $n_{inc}^{-}$ it  oscillates and after the departure, an exponential decay sets in. Using the maximal group and detector velocities, these incidence times are:
\begin{equation}
    n_{inc}^{\pm}=\frac{|\xi|}{1\pm 2\gamma\tau}
    \label{eq: ncpm}
\end{equation}
Note that the dwell time  between the peaks, $\Delta n:=n_{inc}^{-}-n_{inc}^{+}=|\xi|(4\gamma\tau/(1-4(\gamma\tau)^2))$,  increases as a function of $\tau$. The incidence times in the simulation result in Fig. \ref{fig: PdetArrival}a agree with this last equation. 

In the opposite case, when $\xi>0$,  the detector starts to the right of the particle and moves away at a constant velocity. For $\tau$ less than the critical value, the detector outruns the particle and $P_\textrm{det}$ is dominated by the first measurement and so is exponentially small in $\xi$. We get
\begin{equation}
    P_\textrm{det} \approx J_{1+\xi}^2(2\gamma\tau)
    \sim \frac{(\gamma\tau)^{2+2\xi}}{(\Gamma(\xi+1))^2}
    \label{eq: rescaling}
\end{equation}
For $\tau$ larger than the critical value, the particle eventually catches up to and passes the slowly moving detector.  In this case, $F_n$ is maximal at the time of meeting, and falls off like a power-law for large $n$, and the dependence on $\xi$ is only in the phase, as in the case of the SMP.
\begin{figure}[ht]
\includegraphics[width=9.5cm, height=5cm]{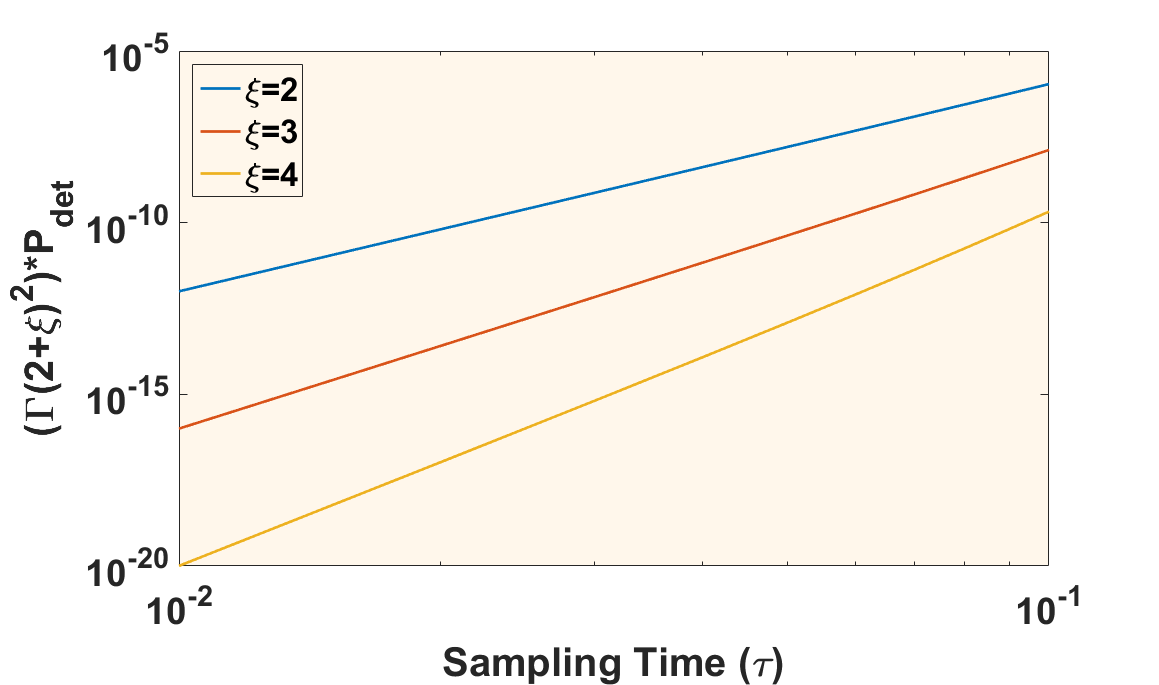}
\centering
\caption{The total detection probability, scaled by $\Gamma^2(2+\xi)$, as a function of $\tau$ in the case of the infinite line with the RMP for several values of positive $\xi$ ($\xi=2,3,4$). Note that the exponent of the power-law behavior ($(\Gamma(2+\xi))^2 P_{det}=\tau^{2+2\xi}$) agrees with Eq. \eqref{eq: rescaling}} 
\label{rescaling}
\end{figure}

 \begin{figure*}[t]
    \centering
    \includegraphics[width=17cm, height= 5cm]{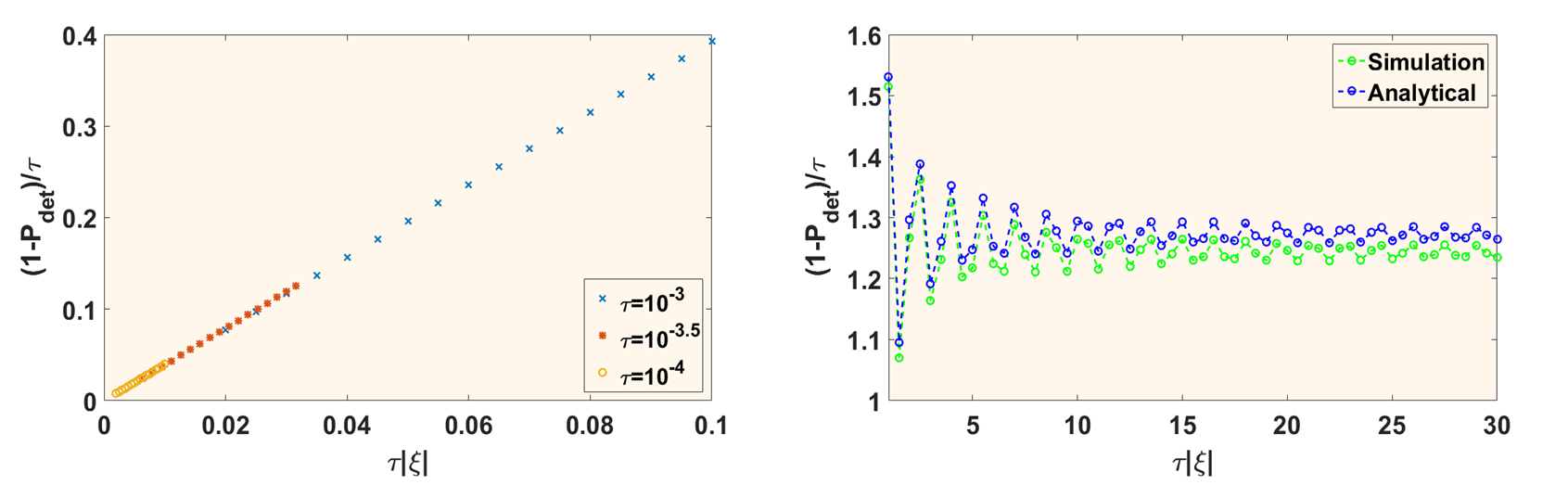}
    \caption{The survival probability $S_{\infty}=1-P_{det}$ for the RMP arrival problem on an infinite line with negative $\xi$. The left figure shows the limit of small $|\xi|$, where the survival probability is close to linear, this agree with Eq. \eqref{eq: Sinflin}. The results were obtained with several sampling time values ($\tau=10^{-4},10^{-3.5},10^{-3}$). The right figure presents the agreement between the simulation and Eq. \eqref{eq: Sinf}, note the re-scaling of axis. In addition, the limit at large $|\xi|\tau$ is seen to agree with our prediction and the oscillation and power-law convergence around this value is apparent. In the simulation of both figures, we take $\gamma=1$ and the cutoff of infinite sum of total detection probability is $N=1000$. To achieve negligible error in simulating Eq. \eqref{eq: Sinf}, the cutoff ($K$) needs to be greater than the argument $2|\xi|\tau$; we chose $K=300$ in the presented calculation.  }
    \label{fig: PdetNegative}
\end{figure*}

\subsection{Zeno effect in the RMP\label{moveZeno}}

The quantum Zeno effect (also known as the Turing paradox) is a feature of quantum mechanical systems allowing a particle's time evolution to be arrested by measuring it frequently enough \cite{Zeffect0,Zeffect1,itano1990quantum}.
 In the quantum first-detection problem, it is manifested as the sampling time $\tau$ goes to zero. For example, when the detection site is fixed in the tight-binding model, as discussed in Ref.~\cite{Harel}, there are two types of Zeno effect. The first is in the arrival problem, where, as the detection sampling period approaches 0, the total detection probability goes to 0. The second case is in the return  problem, in which case, $P_{det}\rightarrow 1$ when $\tau\rightarrow 0$.      
\\\\In the RMP on the infinite line, where the detector site moves right after every detection attempt,  the physics of the small $\tau$ limit is  very different, depending essentially on the sign of $\xi$, and  the Zeno effect is irrelevant. As noted above, for $\xi>0$, the detector runs away very quickly from the particle and the detection probability is extremely small. For $\xi<0$, on the other hand, for small $\tau$ the detector quickly arrives at the initial particle position before the particle has had any chance to move away, and the  total detection probability  goes to one. Thus, in this case, the RMP manages to completely defeat the Zeno-induced problems of the small-$\tau$ limit of the SMP.

Only for $\xi$ very large, on order $1/\tau$, does the particle have a chance to spread out significantly, so in this regime things become more complicated.
For small sampling time $\tau$ and large $|\xi|$, with $\tau|\xi|$ order 1,  we find (see  Appendix C for the derivation) the total probability to detect the particle is still close to one. The survival probability in this limit is given by:
\begin{equation}
    S_{\infty}=1-P_{det}=2\gamma\tau x \big[J_0^2(x) +  J_1^2(x)\big]
    \label{eq: Sinf}
\end{equation}
where we have defined $x\equiv 2\tau\gamma|\xi|$. Fig.~\ref{fig: PdetNegative} presents two limits of the last equation. The first limit is where $x$ is small so \eqref{eq: Sinf} becomes:
\begin{equation}
    S_{\infty}\sim4\gamma^2\tau^2|\xi| .
    \label{eq: Sinflin}
\end{equation}
When $x$
 is very large, the survival probability approaches the finite limit $4\gamma\tau/\pi$ with algebraically decaying oscillations. More specifically,
 \begin{equation}
    S_{\infty}\sim \frac{4\gamma\tau}{\pi}\left[ 1 - \frac{\cos 2x}{2x}\right] 
\end{equation}
 Fig.~\ref{fig: PdetNegative} show agreement of the simulation and Eq.~\eqref{eq: Sinf} for the finite $x$ case and Eq.~\eqref{eq: Sinflin} for the  small $x$ limit.

\subsection{Large Rings and the Infinite Line}
In Ref.~\cite{Harel}, the behavior of $F_n$ for the SMP for a large but finite ring was discussed.  It was noted that $F_n$ for the large ring closely followed that for the infinite line, up to some critical $n_c$, but the details were not analyzed. In the case of the RMP, the
situation is similar. Physically, the cyclic boundary condition causes  the detector to see the opposite peak of the wave function at this critical $n_c$ and therefore in this region  $F_n$ starts increasing as a power-law. This behavior is similar to behavior for not too large $n$ in the SMP arrival problem which was discussed in Ref.~\cite{infinite_line}. As there, we can find the value of  $n_c$ via considerations of group velocity. The maximum group velocity of the return peaks is $2\gamma$, while the detector's velocity is $1/\tau$, so, for the return problem, their first meeting will occur at: 
\begin{equation}
    n_c=\frac{L}{2\gamma\tau+1}
    \label{eq: nc}
\end{equation}
The cyclic boundary condition means that the free propagation (the direct term in the renewal equation) of the particle on the finite ring can be expressed as a summation over ``image particles" of free propagation on the infinite line. Specifically, as shown in Appendix A,
\begin{align}
    u_n^\textit{Ring}(|0\rangle\rightarrow|0\rangle)&=\sum_{m=-\infty}^{\infty} u_n^\textit{Infinite-Line}(|0\rangle\rightarrow|mL\rangle) \nonumber\\
    &= i^m J_{mL+n}(2n\gamma\tau)
    \label{eq: RILR}
\end{align}
 For $n\ll n_c$, only the $m=0$ term in the sum is not negligible so in this time frame the particle propagates similarly to a particle on an infinite line, and so  is the case for the first-dectection probability  $F_n$.

\begin{figure*}[t]
    \centering
    \includegraphics[width=17cm, height= 5cm]{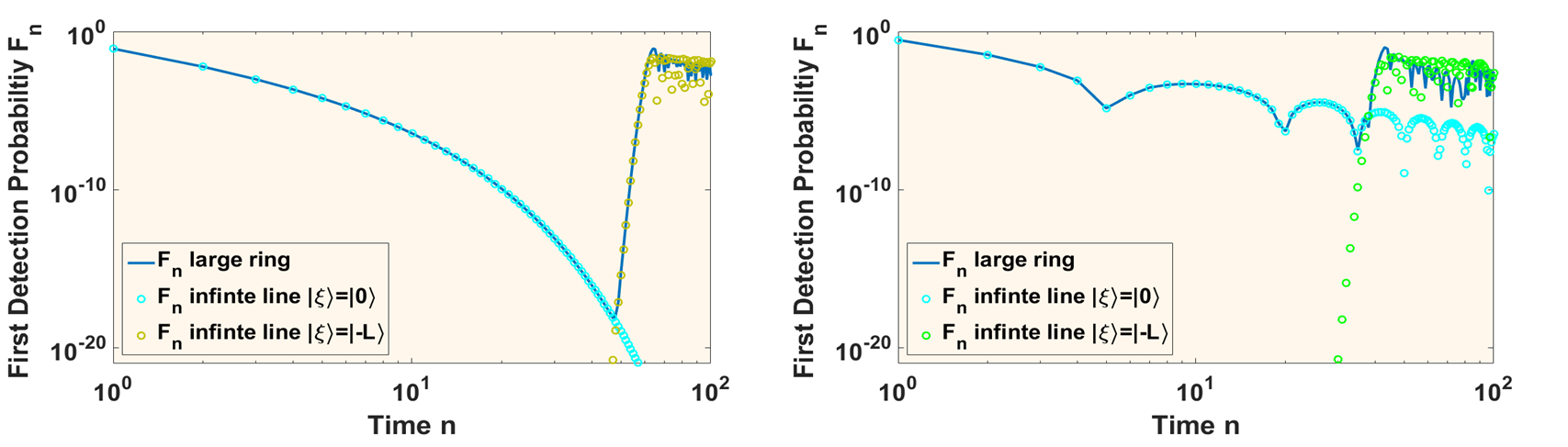}
    \caption{The first-detection probability $F_n$ on a large but finite ring (blue line). In addition, for comparison, we show the $F_n$ of the return problem (cyan circles) and the arrival problem (green circles) with initial detection state $|\psi_d\rangle=|-L\rangle$. Right figure: we used L=100 and sampling time $\tau=0.7$, so the critical  $n_c\approx42$. For $n<n_c$, the oscillation superimposed on  the power-law decay is clearly obtained, which coincides with the behavior of  the slow-moving detector case on an infinite line, together with a monotonic increase in the vicinity of $n_c$. Left figure:  $L=100$ and $\tau=0.3$, so  $n_c\approx63$. For $n<n_c$, the exponential decrease is clearly seen, which coincides with the behavior of the infinite line with a fast-moving detector,  and likewise in the right figure a monotonic increase of $F_n$ in the vicinity of $n_c$ is seen. In both figures, there is  agreement both with Eq.\eqref{eq: nc} and with the qualitative descriptions given in the main text.}
    \label{fig: L large}
\end{figure*}

As we approach the critical $n_c$, the $m=0$ term of the sum  decreases and the $m=\pm 1$ terms become more significant, with the $m=-1$ dominating for $\gamma\tau<1/2$.   Note that the unsuccessful measurement reaction terms of
the renewal equation for $\phi_n$ are negligible  because of the tiny values of $u_n^\textit{Infinite-Line}(|0\rangle\rightarrow|\pm L\rangle)$; a similar phenomenon for the SMP case was discussed in Ref.~\cite{infinite_line}. An approximation of direct terms at this time gives the power-law increase. In Fig.~\ref{fig: L large}, we simulate $F_n$ on a $L=100$ ring  for both the fast and slow detector cases. In addition, the simulation of $F_n$ on the infinite line of the return and arrival amplitude are shown. The match  between the  first-detection probability on the finite ring with that on the infinite line before and near $n_c$ and the subsequent agreement with Eq. \eqref{eq: nc} are shown in Fig. \ref{fig: L large}. 

\section{Summary and Discussion}
We investigated the Running Measurement Protocol on the tight-binding quantum walk model under repeated stroboscopic measurement and mapped it to an equivalent stationary detector system with a modified unitary evolution operator. We computed the first-detection probability distribution $F_n$ and other statistical quantifiers of the process like $\langle n\rangle$ and $\textit{Var}(n)$ for a ring and $F_n$ for an infinite line. For the ring, the modified unitary evolution operator breaks the reflection symmetry and restores perfect detection for the arrival problem. For the return problem, it increases   $\langle n\rangle$ while reducing $\textit{Var}(n)$ close to the small $\tau$ Zeno limit, in line with the findings of Ref.~\cite{sinkovicz2015quantized}. This fact is important, since the average time for first detection is $\tau\langle n\rangle$ which is very small but now the variance is also  small, so that  the Zeno effect is eliminated. The increase of  $\langle n \rangle$ in the  return problem due to the elimination of quantum interference is observed also in open quantum systems when  decoherence effects are added~\cite{sinkovicz2015quantized}.

For the infinite line, one major new feature of the RMP is the presence of a critical detector velocity.  If $\gamma\tau<0.5$, the detector outruns the particle, so if $\xi> 0$, the total detection probability is very small and goes to zero as $\xi$ increases. For $\gamma\tau>0.5$, the detector is slow enough that it captures a range of particle velocities, and $P_\textit{det}$ is significantly larger. Hence we have a simple quantum dynamical phase transition. This behavior is seen in the numerical results in Fig. \ref{fig: PdetArrival}. This transition obtains also in a biased discrete quantum walk \cite{vstefavnak2009recurrence}. The second transition, at $\gamma\tau=2.32$, at which oscillations in $P_\textit{det}$ set in for $\xi>0$, is analogous to the transition seen at $\gamma\tau=\pi/2$ for the corresponding stationary detector case.
\\\\The ability to tune, by varying $\tau$ and hence the detector velocity, which range of group velocities of the particle the detector will encounter, allows new behaviors to emerge. In the case where the detector outruns all possible velocities of the particle, an exponential decay of the detection probability with attempt number  emerges (see Fig. \ref{fig: Fn&trajectory}). When the detector velocity is such that it encounters particles within a certain range of group velocities, one sees  behavior similar to that of the stationary protocol with power-law decay and recurrence of the system. In the critical case ($\gamma\tau=0.5$) where only the maximal velocity is probed,  there is a power-law behavior with an anomalous exponent (Eq. \eqref{eq: FnGT0.5}).  A similar special power-law decay at the border between exponential  and power-law decay of the arrival probability is observed in Brownian motion with a  boundary with a velocity that decays as a power-law in time~\cite{krapivsky1996life}.  
\\\\ The moving frame in addition breaks the left/right symmetry between $\pm\xi$ in the total detection probability. For $\xi>0$, the total probability to detect the particle decays exponentially.  For  large values of negative $\xi$, it converges to a finite value (which decreases with increasing $\tau$) as $
\xi\to-\infty$. The exponential decay of the total probability to detect the particle as a function of the initial   particle-detector distance is found also in Brownian motion with escape through ballistically moving  boundaries \cite{krapivsky1996life,bray2007survival}.  

 A possible generalization of RMP is to move the detector more than one step between measurements. This would allow independent control over the measurement interval $\tau$  and the detector velocity. One complication of this scheme is the possibility of hopping over the particle and missing it completely. To counter this, one would have to widen the detector to compensate.  It would be interesting to pursue this line of inquiry.

 Our results can, in principle, be tested experimentally in a cold atom system, given the possibility of realizing a conveyor belt of cold atoms \cite{Kuhr,Schrader}  with a very local detection of a single site in an optical lattice. This conveyor belt protocol is equivalent to the Running Measurement Protocol and its dynamics are again representable as a  modified unitary propagator a la Eq. \eqref{eq: Modifated Propagator}. 
\\\\The support of Israel Science Foundation's grant 1898/17 is acknowledged. 

\appendix
\section{Poisson summation}
In this appendix we will discuss the finite version of Poisson summation and its application to the finite ring propagator as discussed in main text. Let $f(x)$ a complex-valued function of a real variable $x$, periodic  with period $L_0$. Therefore it can expressed as a Fourier series:
\begin{equation}
    f(x)=\sum_{l=-\infty}^{\infty} c_l e^{-i\frac{2\pi lx}{L_0}}
    \label{eq:four}
\end{equation}
where the Fourier coefficients  are given by:
\begin{equation}
    c_l=\frac{1}{L_0}\int_{-L_0/2}^{L_0/2} f(x) e^{i\frac{2\pi lx}{L_0}} dx
\end{equation}
Plugging \eqref{eq:four} into the Riemann sum of $f$ over a period, the only contributions that survive are for $l=mN$, for integer $m$, and we get
the finite version of the Poisson summation forumula\cite{trefethen2014exponentially}:
\begin{equation}
    \sum_{k=0}^{N-1} f\left(\frac{L_0 k}{N}\right)=\sum_{m=-\infty}^{\infty} c_{mN}
    \label{eq: PoissonSum}
\end{equation}
We can use this formula to calculate the ($|0\rangle\rightarrow|\xi\rangle$) propagator represented  as a sum over the wavenumber $k$ of the function $f(k)=e^{ik\xi}e^{-iE_kn\tau}$. Note that $f(k)$  is continuous and periodic with period $2\pi$. In addition, the number of sites ($L$) is equal to $N$ in Eq.\eqref{eq: PoissonSum}. Therefore the Fourier coefficients are: 
\begin{align}
     c_{mL}&=\frac{1}{2\pi} \int_{-\pi}^{\pi} e^{ik\xi}e^{ikn+2in\gamma\tau\cos k} e^{imLk} dk\nonumber \\
     \nonumber\\
     &= (i)^{n+mL+\xi}J_{|n+mL+\xi|}(2\gamma\tau n)\nonumber\\
     &=\langle mL+\xi|\Hat{U}(n\tau)|0\rangle
\end{align}
This, together with Eq.~\eqref{eq: PoissonSum}, gives Eq.~\eqref{eq: RILR} in the main text. 

\section{The behavior of $u_n$ in the RMP for the infinite line with $\gamma\tau=0.5$ }
In this Appendix, we analyze the return amplitudes for the RMP infinite line problem at the critical $\gamma\tau=0.5$ from the dispersion relation $E(k)$ using the Fourier-Tauber theorem which is discussed in Ref.~\cite{spectral_dim}. The goal is to calculate the exponent associated with the power-law decay from the behavior   of the measured spectral density of states (MSDOS, see Ref.  \cite{spectral_dim} for the definition) near the singular point using the Mellin transform. The exponent is determined by the location of the pole with the largest real part. The novel aspect of the case treated in this appendix is that the effective dispersion relation for $E(k)\equiv \lambda(k)/\tau$ has an inflection point at $k^\star=-\pi/2$, and so we have  to expand to 3rd order in Taylor series.  The Mellin transform of the return amplitude MDOS ($f(E)$) is:   
\begin{equation}
    M^{+}[f,s]=\frac{1}{2\pi}\int_{-\pi}^{\pi} dk (E(k)-E^\star)^{s-1} \theta(E(k)-E^\star)
\end{equation}
where $\theta(\cdot)$ is the Heaviside step function. Note that the integral includes only those energies  greater than the singular energy so from this equation we will get only $M^{+}$. Equivalently, it is possible to calculate $M^{-}$ from the energies lower than $E^*$ because the expansion is around an inflection point, as opposed to the case of an extremal point, and so both sides of the energies contribute.  Expanding $E(k)$ to 3rd order around $k^\star$ we get: 
\begin{equation}
     M^{+}[f,s]=\frac{1}{2\pi}\int_{k^\star}^{k^\star+\delta} dk (\frac{1}{6} E^{(3)}(k^\star)(k-k_1)^3)^{s-1} 
\end{equation}
where $\delta\ll 1$, since the falloff is controlled by the region near the critical point.  The $\theta$ function  causes  asymmetric limits of integral around $k^\star$. Changing variables to $x:=k-k^\star$, we get: 
\begin{equation}
    M^{+}[f,s]=C\int_{0}^\delta x^{3s-3}
\end{equation}
$C$ is defined as the other terms which do not depend on $x$. The integral yields a one first-order pole at $s=2/3$. The next step is to find $f(E^\star+\epsilon)$. This is done via an inverse Mellin transform.    
\begin{equation}
    f(E^\star+\epsilon)=\frac{1}{2\pi i }\int_{c-i\infty}^{c+i\infty} ds\epsilon^{-s} C\frac{\delta^{3s-2}}{3s-2}
\end{equation}
The notation implies this is a line integral taken over a vertical line in the complex plane. Using the residue theorem,
we find the power-law of $f(E)$ around $E^\star$ is $-2/3$. According the Fourier-Tauber theorem, the  large-$n$ decay of $u_n$ is then a power-law with exponent $-1/3$. This result matches with the power-law of $J_n(n)$ we saw in Eq. \eqref{eq: AS 1}, the difference between $u_n$ and  Eq. \eqref{eq: AS 1} is only the phase of $e^{-i\pi n/2}$ which does not change the decay exponent.

\section{Derivation of $P_\textit{det}$ for the infinite line arrival problem}
As mentioned in the main text, we can solve for $P_\textit{det}(\xi,\tau)$ in the small $\tau$ limit.  For $\xi>0$, $P_\textit{det}(\xi,\tau)$ falls off very rapidly with $\xi$, and is dominated by the ``direct" term, $|\langle \xi|U(n)|0 \rangle|^2$, as the denominator in $\phi(z)$ is $1$ to leading order in $\tau$, giving
\begin{equation}
    F_n \approx J_{|n+\xi|}^2(2\gamma\tau n) \approx \frac{(\gamma\tau n)^{2n+2\xi}}{((n+\xi)!)^2}
\end{equation}
$P_\textit{det}$ for $\xi\ge 0$ is then dominated for small $\tau$ by $F_1$, yielding 
\begin{equation}
    P_\textit{det}  \approx \frac{(\gamma\tau )^{2\xi+2}}{((\xi+1)!)^2}
\end{equation}
The situation is very different for $\xi<0$.  Again, the direct term dominates $F_n$, but the dominant term is
$n=-\xi$, where $F_{|\xi|}\approx J_0^2(2\gamma\tau|\xi|)\approx 1$.  To next order in $\tau^2$, $F_{|\xi|\pm 1}$ also contribute, yielding an additional $J_1^2(2\gamma\tau(|\xi|+1))+J_1^2(2\gamma\tau(|\xi|-1))$ to $P_\textit{det}$. 
Whereas for fixed $\xi$, this term is down by a factor $\tau^2$, for large $\xi \sim {\cal{O}}(1/\tau)$, this term can be of order 1.  Summing up all the direct contributions, we have
\begin{align}
    P_\textit{det}^\textbf{direct} &= J_0^2(2\gamma\tau |\xi|) + \sum_{k=1}^\infty \left[J_k^2(2\gamma\tau (|\xi|+k)) + J_k^2(2\gamma\tau (|\xi|-k))\right] \nonumber\\
    &\approx J_0^2(2\gamma\tau |\xi|) + 
    2\sum_{k=1}^\infty J_k^2(2\gamma\tau|\xi|) \nonumber\\
    & = 1
    \label{Pdetdir}
\end{align}
where we have used the identity 9.1.76 from Ref. \cite{abramowitz1964handbook} in the last line.
This result gives $P_\textit{det}$ correctly to leading order in the small $\tau$, finite $\tau\xi$ limit. 

To next order, we have to include the first correction from the denominator in $\phi(z)$. We have
\begin{equation}
    u(z) \approx 1 + izJ_1(2\gamma\tau) \approx 1 + i\gamma\tau z .
\end{equation}
The numerator is as before, so
\begin{equation}
    \varphi(z)= \frac{\sum_n (iz)^n J_{|n+\xi|}(2n\tau)}{1+i\gamma\tau z}
\end{equation}
so that
\begin{equation}
    \varphi_n \approx (i)^n\left[J_{|n+\xi|}(2n\gamma\tau) - J_{|n-1+\xi|}(2n\gamma\tau)\right]
\end{equation}
The sum, is as above, dominated by the $n=|\xi|$ term, so that, to this order, we have
\begin{equation}
    P_\textit{det}\approx \sum_{k=0}^\infty J_k^2(2\gamma\tau|\xi|) - 4\gamma\tau J_k(2\gamma\tau|\xi|) J_{k+1}(2\gamma\tau|\xi|)
\end{equation}
We thus see that $P_\textit{det}\approx 1 - \gamma\tau{\cal{G}}(2\gamma\tau|\xi|)$. The function ${\cal{G}}$ is
given by
\begin{subequations}
\begin{align}
    {\cal{G}}(x)&=4\sum_{k=0}^\infty J_k(x) J_{k+1}(x)\label{eq:G1}\\
                &=2x\left[J_0^2(x) + J_1^2(x)\right]\label{eq:G2}
\end{align}
\end{subequations}
This last identity can be proven by differentiating Eq. \eqref{eq:G1} and seeing that the sum telescopes:
\begin{align}
    {\cal{G}}' &= 4\sum_{k=0}^\infty [J_k' J_{k+1} + J_k J_{k+1}'] \nonumber\\
    &=2\sum_{k=0}^\infty [J_{k-1}-J_{k+1}]J_{k+1} + J_k[J_k - J_{k+2}]
    \nonumber\\
    &= 2 [ J_{-1}J_1 + J_0^2] + 2\sum_{k=0}^\infty [J_k J_{k+2}-J_k^2] \nonumber\\
    &\qquad {}+ 2\sum_{k=0}^\infty [J_k^2 - J_k J_{k+2}]\nonumber\\
    &= 2\left[J_0^2 - J_1^2\right]
\end{align}
Similarly, differentiating Eq. \eqref{eq:G2} gives the same result.
For small $x$, ${\cal{G}}(x)$ behaves linearly, ${\cal{G}}(x)\approx 2x$
The large-$x$ asymptotics of ${\cal{G}}(x)$ are easy to calculate:
\begin{equation}
    {\cal{G}}(x) \approx \frac{4}{\pi}\left[1 - \frac{\cos 2x}{2x}\right]
\end{equation}

In principle, we also have to include the correction induced by the $\pm k$ in the argument of the Bessel functions in Eq. (\ref{Pdetdir}).  However, this correction, being odd in $k$, does not give a net contribution to $P_\textit{det}$.

\bibliographystyle{iopart-num}
\bibliography{./bibliography2} 
\iffalse

\fi

\end{document}